# Quantum Metric Unveils Defect Freezing in Non-Hermitian Systems

Karin Sim[1,*], Nicolò Defenu[1], Paolo Molignini[2,3], and R. Chitra[1]

[1]*Institute for Theoretical Physics, ETH Zürich, 8093 Zurich, Switzerland*
[2]*Cavendish Laboratory, University of Cambridge, 19 J J Thomson Avenue, Cambridge CB3 0HE, United Kingdom*
[3]*Department of Physics, Stockholm University, AlbaNova University Center, 106 91 Stockholm, Sweden*

Non-Hermiticity in quantum Hamiltonians leads to nonunitary time evolution and possibly complex energy eigenvalues, which can lead to a rich phenomenology with no Hermitian counterpart. In this work, we study the dynamics of an exactly solvable non-Hermitian system, hosting both $\mathcal{PT}$-symmetric and $\mathcal{PT}$-broken modes subject to a linear quench. Employing a fully consistent framework, in which the Hilbert space is endowed with a nontrivial dynamical metric, we analyze the dynamics of the generated defects. In contrast to Hermitian systems, our study reveals that $\mathcal{PT}$-broken time evolution leads to defect freezing and hence the violation of adiabaticity. This physics necessitates the so-called metric framework, as it is missed by the oft used approach of normalizing quantities by the time-dependent norm of the state. Our results are relevant for a wide class of experimental systems.

*Introduction.*—Non-Hermitian Hamiltonians [1,2] provide a framework to explore a complex array of out-of-equilibrium phenomena. Far from being a purely mathematical pursuit, non-Hermitian descriptions have been employed widely in both classical and quantum systems. The most well-known examples include the study of non-Hermitian spin chains in the context of the Kardar-Parisi-Zhang equation [3], localization of particles in an imaginary vector potential to explain vortex depinning [4], and open quantum systems in the no-jump limit [5]. Non-Hermiticity has unveiled a plethora of interesting phenomena, such as quantum phase transitions without a gap closure [6], anomalous behaviors of quantum emitters [7,8], tachyonic physics [9,10], and unconventional topology [11–13], to name a few. The interest in non-Hermitian systems is further enchanced by the concomitant experimental realizations in diverse platforms: optical [14], semiconductor microcavities [15], and acoustic [16]. Besides, non-Hermitian Hamiltonians can also be directly engineered in a fully controllable manner using conventional quantum gates via Naimark dilations [17–19].

Non-Hermitian Hamiltonians which preserve $\mathcal{PT}$ symmetry (i.e., the combined operation of parity and time reversal) [20] constitute a special class of systems possessing a real spectrum, prompting their interpretation as a natural extension to conventional quantum mechanics [21].

When $\mathcal{PT}$ symmetry is spontaneously broken, exceptional points (EPs) [22] arise, signaling the coalesence of eigenvectors and the emergence of complex eigenvalues. EPs have been a subject of much recent attention, both theoretically [2,23] and experimentally [24].

Non-Hermiticity ushers in new challenges to fundamental concepts in conventional quantum mechanics, necessitating a more general framework. Foremost is biorthogonal quantum mechanics [25], which has been widely studied in the context of $\mathcal{PT}$-symmetric Hamiltonians, but has its limitations. More often, the time-dependent Schrödinger equation is directly used in conjunction with an *ad hoc* explicit normalization of time-dependent probabilities and observables, an approach ubiquitous in the open quantum systems community [26–31]. As we shall see in this work, this method can fail to capture salient aspects of the physics. A more robust and consistent formulation of non-Hermitian quantum mechanics is provided by the so-called metric framework, wherein the Hilbert space is nonstationary and endowed with a nontrivial time-dependent "metric" [32–34]. It can be regarded as a generalization of biorthogonal quantum mechanics [25], encompassing spontaneous $\mathcal{PT}$-broken scenarios as well. This framework is vital to recover fundamental theorems of quantum information [35], as well as being especially relevant for the quantum brachistochrone problem [36] and the evolution of entanglement [37]. From the fundamental perspective, this approach has several advantages: the norm of the wave function is conserved, which implies that the notion of probability and the values of the observables remain physical at all times.

In the Hermitian realm, quantum quenches and driving have emerged as tools of choice to explore the rich array of dynamical phenomena [38–41]. The study of analogous dynamics in non-Hermitian systems has also garnered massive attention recently [28–30,42]. In this work, we consider the famous example of Kibble-Zurek (KZ) scaling which dictates how the density of topological defects scales when a coupling is quenched across a quantum critical point [43]. The KZ scaling predicts that the defect density scales as a power law with quench time, where the exponents are determined by the static critical exponents [44]. Using the wave function normalization approach, recent works predicted a modified KZ scaling when a system is quenched across EPs [27,45], thereby recovering adiabaticity. On the other hand, the breakdown of adiabaticity was seen experimentally in dissipative superconducting qubits governed by effective non-Hermitian Hamiltonians [46]. This behooves the question of whether a more consistent approach is required to capture the violation of quantum adiabaticity. In this Letter, we rigorously investigate this fundamental question using an exactly solvable non-Hermitian model. We show that the metric plays a crucial role in the violation of quantum adiabaticity when EPs are traversed adiabatically.

*Metric framework.*—We consider a dynamical Hilbert space endowed with a positive-definite operator $\rho(t)$, which is time dependent in general and appears as a weight factor in the inner product of this Hilbert space as $\langle \cdot, \cdot \rangle_{\rho(t)} := \langle \cdot | \rho(t) \cdot \rangle$ [32–34]. In this Letter, we employ the terminology "metric" to refer to $\rho(t)$ because it has become established in the literature [21,32,33,37]. However, we note that $\rho(t)$ is not a metric in the strict sense of a map in a metric space, and it does not correspond to the quantum geometric tensor discussed in Refs. [47,48]. The dynamics of the Hilbert space $\mathcal{H}_{\rho(t)}$ is encoded in the time evolution of $\rho(t)$, given by [33]

$$i\dot{\rho}(t) = H^\dagger(t)\rho(t) - \rho(t)H(t), \quad (1)$$

where the overdot denotes time derivative. Provided that a solution to Eq. (1) can be found [33], we can map the system to a Hermitian Hamiltonian $h(t) = \eta(t)H(t)\eta^{-1}(t) + i\dot{\eta}(t)\eta^{-1}(t)$, where we have introduced the square-root decomposition $\rho(t) = \eta^\dagger(t)\eta(t)$. The Hamiltonian $h(t)$ acts in a different Hilbert space $\mathcal{H}$ [33], where the non-Hermiticity is encoded in the dynamics of $\eta(t)$. The explicit derivation of the metric $\rho(t)$ in various physical situations has been the subject of several investigations [34,49–51]. A recent study explicitly computed $\eta(t)$ for a two-level system akin to ours [52]. Yet, the exact form of the metric in the context of infinite-dimensional Hilbert spaces has only been obtained for certain solvable models [53], while its existence in generic many-body systems remains an open question.

Time evolutions in the Hilbert spaces $\mathcal{H}_{\rho(t)}$ and $\mathcal{H}$ are generated by the respective Hamiltonians, $H(t)$ and $h(t)$, via the time-dependent Schrödinger equation (TDSE)

$$i\frac{d}{dt}|\psi(t)\rangle = H(t)|\psi(t)\rangle$$
$$i\frac{d}{dt}|\Psi(t)\rangle = h(t)|\Psi(t)\rangle, \quad (2)$$

where the states are related by $|\Psi(t)\rangle = \eta(t)|\psi(t)\rangle$. The construction of $\rho(t)$ guarantees that $h(t)$ is Hermitian, such that the unitarity of the time evolution is restored, giving $\langle\psi(t)|\rho(t)|\psi(t)\rangle = \langle\Psi(t)|\Psi(t)\rangle = 1$ at all times $t$ [33]. On the other hand, the expectation value of an operator $\hat{o}: \mathcal{H} \to \mathcal{H}$ is given by

$$\langle O(t)\rangle_{\text{metric}} = \langle\Psi(t)|\hat{o}|\Psi(t)\rangle = \langle\psi(t)|\rho(t)\hat{O}(t)|\psi(t)\rangle, \quad (3)$$

where $\hat{O}(t): \mathcal{H}_{\rho(t)} \to \mathcal{H}_{\rho(t)}$ is defined as $\hat{O}(t) = \eta^{-1}(t)\hat{o}\eta(t)$. Therefore, given an operator $\hat{o}$ and state $|\Psi(t)\rangle$ describing an observable in $\mathcal{H}$, the operator $\hat{O}(t)$ and state $|\psi(t)\rangle$ in $\mathcal{H}_{\rho(t)}$ describe the same observable of the system [33]. In other words, Eq. (3) describes a physically meaningful time-dependent expectation value which is consistent across both representations, justifying the probabilistic interpretation of quantum mechanics. In contrast, the expectation of $\hat{o}$ calculated from a simple normalization by the time-dependent norm is given by

$$\langle O(t)\rangle_{\text{norm}} = \frac{\langle\psi(t)|\hat{o}|\psi(t)\rangle}{\langle\psi(t)|\psi(t)\rangle} \quad (4)$$

as was done, for example, in Refs. [26–31].

It is worth noting that though Eq. (1) can have an infinite family of solutions in general, a unique $\rho(t)$ can be determined by providing a specific initial condition. Additionally, as the non-Hermitian contribution to the Hamiltonian vanishes, $\rho(t) \to \mathbb{1}$, and the Hilbert spaces $\mathcal{H}_{\rho(t)}$ and $\mathcal{H}$ should coincide. Then, a natural choice of $\eta(t)$ shall tend to the identity operator in this limit, such that $h(t) = H(t)$ when the non-Hermiticity vanishes. All other square roots are related to this by unitary transformations corresponding to rotations.

*Exactly solvable model.*—To highlight the nontrivial role played by the metric, we consider an exactly solvable model of effective two-level systems parameterized by momentum $k$. This is given by the Hamiltonian [54]

$$H_k(t) = k\sigma_x + i\gamma\sigma_y + Ft\sigma_z \quad (5)$$

given in natural units $\hbar = c = 1$, where $\sigma_i$ denotes the Pauli matrices and $F, k, \gamma \in \mathbb{R}$. Equation (5) is a generalization of the Hamiltonian presented in Ref. [55] and realized experimentally in Ref. [56], by adding a real drive term $Ft$ and applying a basis rotation. There, $\gamma$ corresponds to the imaginary tachyon mass [55], $k$ is the momentum, and $F$ is a force [57]. The dimensionless term $(\gamma^2/F)$ sets the scale for the extent of non-Hermiticity in our model. $\mathcal{PT}$ symmetry is realized in our model by the operators $\mathcal{P} = \sigma_y$ and $\mathcal{T} = -i\sigma_y \mathcal{K}$ where $\mathcal{K}$ is complex conjugation, such that $[H_k, \mathcal{PT}] = 0$. At the EP, the spontaneous breaking of this symmetry occurs and the states are no longer eigenstates of the $\mathcal{PT}$ operator. The instantaneous eigenvalues of Eq. (5) are given by $E_{\pm,k}(t) = \pm\sqrt{F^2 t^2 + k^2 - \gamma^2}$, as shown in Fig. 1. By tuning the momentum $k$ and the imaginary mass $\gamma$, our Hamiltonian permits us to study the evolution of two different types of modes: those that undergo fully $\mathcal{PT}$-symmetric evolution, $|k| \geq |\gamma|$ and those that pass through the EPs during their evolution, $|k| < |\gamma|$.

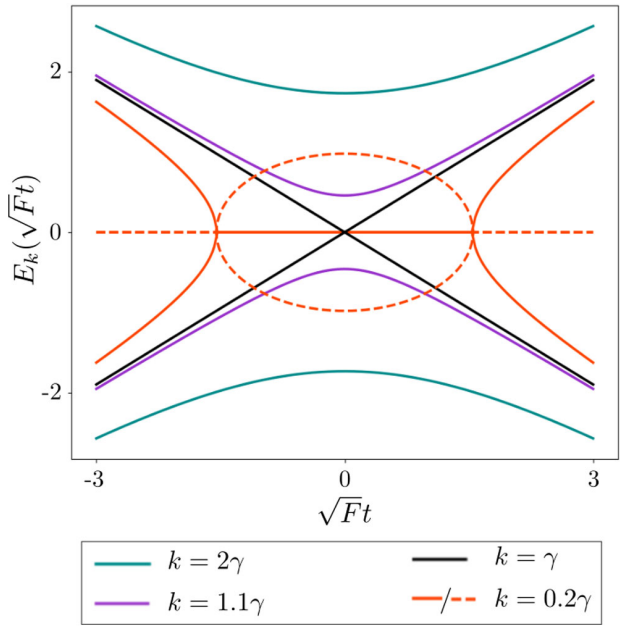

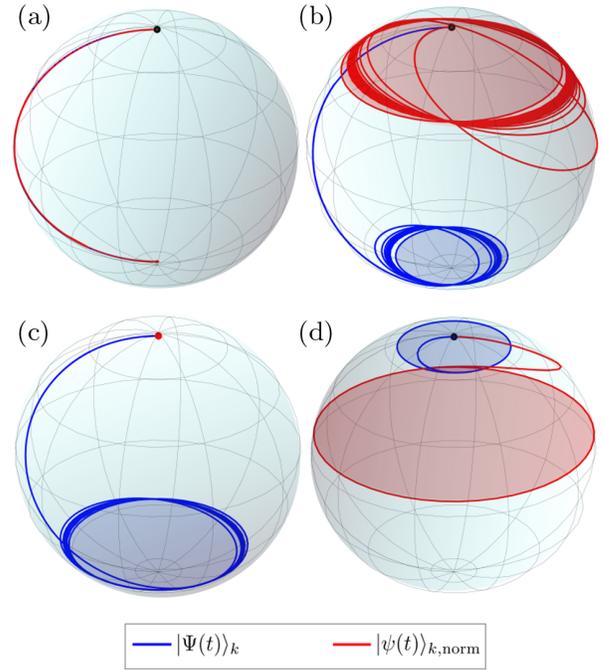

FIG. 1. The instantaneous spectrum of the non-Hermitian Hamiltonian [Eq. (5)] as a function of time, where $\gamma = 1$ and $(\gamma^2/F) = 2.5$. While the static system only has exceptional points (EPs) at $k = \pm\gamma$, the EPs are also found for $|k| < |\gamma|$ in the dynamical case. For $k = 0.2\gamma$, the solid and dashed lines show the real and imaginary parts, respectively. Our model allows us to track both $\mathcal{PT}$-broken and $\mathcal{PT}$-symmetric evolution.

FIG. 2. The evolution of the normalized states $|\Psi(t)\rangle_k = \eta_k(t)|\psi(t)\rangle_k$ (blue) and $|\psi(t)\rangle_{k,\text{norm}} \equiv |\psi(t)\rangle_k/\||\psi(t)\rangle_k\|$ (red) on the Bloch sphere for (a) $k = 2\gamma$, (b) $k = 1.1\gamma$, (c) $k = \gamma$, and (d) $k = 0.2\gamma$, c.f. Fig. 1. Here $\gamma = 1$, $(\gamma^2/F) = 2.5$ which is far from the adiabatic limit, and the evolution is between the asymptotic initial state at the north pole (black dot) and a distant end point at $t = (80/\sqrt{F})$. For $k \gg \gamma$, the dynamics of $|\Psi(t)\rangle_k$ and $|\psi(t)\rangle_{k,\text{norm}}$ are in good agreement with each other, see (a). However, this is not true for $k \approx \gamma$ even for a $\mathcal{PT}$-symmetric evolution, see (b). For $k = \gamma$, the dynamics of $|\Psi(t)\rangle_k$ is completely due to the metric, as $|\psi(t)\rangle_{k,\text{norm}}$ stays at the north pole and does not evolve in time, see (c). The discrepancy between $|\Psi(t)\rangle_k$ and $|\psi(t)\rangle_{k,\text{norm}}$ is significant for $\mathcal{PT}$-broken evolution too, see (d).

The dynamics of our model is exactly solvable through Eqs. (1) and (2), making it ideal for illustrating an accurate description of non-Hermitian physics. In analogy to the Hermitian Landau-Zener problem [58], we time evolve the system between the asymptotic limits $t \to \pm\infty$, which correspond to Hermitian initial and end points. Using the exact solution for $\rho_k(t)$ with the Hermitian initial condition $\rho_k(t \to -\infty) = \mathbb{1}$ valid for all $k$, we can map our problem to a Hermitian Hamiltonian $h_k(t)$, where the dynamical richness of $\rho_k(t)$ is directly encoded in the dynamics of $h_k(t)$ [59].

In contrast to the original Hamiltonian $H_k(t)$, we find that $h_k(t)$ does not describe a linear quench, where the extent of its departure from the linear quench regime is dictated by the parameters $(\gamma^2/F)$ and $\delta = [(k^2 - \gamma^2)/2F]$. This modified dynamics due to $\rho_k(t)$ influences the evolution of the state $|\Psi(t)\rangle_k$, defined in Eq. (2), for certain parameter regimes. For $k \gg \gamma$, i.e., very weak non-Hermiticity, this modification is rather insignificant and $|\Psi(t)\rangle_k$ and $|\psi(t)\rangle_{k,\text{norm}} \equiv |\psi(t)\rangle_k/\||\psi(t)\rangle_k\|$ are in good agreement with each other, as shown in Fig. 2(a). However, this equivalence breaks down when $k \sim \gamma$ (even when the $\mathcal{PT}$ symmetry is not broken) and in the $\mathcal{PT}$-broken evolution $|k| < \gamma$, as shown in Figs. 2(b)–2(d). Curiously, for the critical value $k = \gamma$, the evolution of the state $|\Psi(t)\rangle_k$ is entirely due to the dynamics of $\rho_k(t)$. Consequently, the state $|\psi(t)\rangle_{k,\text{norm}}$ stays at the north pole of the Bloch sphere and does not evolve, as shown in Fig. 2(c).

Another striking difference concerns the symmetry of the Hamiltonians. In the Hermitian limit $\gamma \to 0$, we have $\sigma_z H_k(t) \sigma_z = H_{-k}(t)$, from which we deduce the even parity of the $\sigma_z$ expectation value with respect to $k$. This even parity persists in the non-Hermitian case, which we demonstrate via appropriate symmetry transformations considering both the left and the right eigenstates of $H_k(t)$ (see [59]). The spin expectation value calculated using the metric formalism preserves the even parity, as the construction of $\rho_k(t)$ takes into account the states evolved using both $H_k(t)$ and $H_k^\dagger(t)$ [59]. This also ensures the symmetry $\sigma_z h_k(t) \sigma_z = h_{-k}(t)$ in the mapped Hermitian Hamiltonian, consistent with the symmetry analysis [59]. As a consequence, while $|\Psi(t)\rangle_k = |\Psi(t)\rangle_{-k}\ \forall\ k$, this symmetry is not respected by $|\psi(t)\rangle_{k,\text{norm}}$ in general.

*Spin expectation.*—The different state trajectories predicted by the two methods lead to qualitatively different spin expectation values $\langle\sigma_z(t)\rangle_{k,\text{metric}}$ and $\langle\sigma_z(t)\rangle_{k,\text{norm}}$, calculated from Eqs. (3) and (4) by setting $\hat{o} = \sigma_z$ [59]. Much like the time-evolved state $|\psi(t)\rangle_{k,\text{norm}}$, we find that $\langle\sigma_z(t)\rangle_{k,\text{norm}}$ does not have a definite parity in $k$. On the other hand, $\langle\sigma_z(t)\rangle_{k,\text{metric}}$ is even in $k$, consistent with the aforementioned symmetry analysis. The exact results for the spin expectation values in the asymptotic limit $\langle\sigma_z(\infty)\rangle \equiv \langle\sigma_z(t \to \infty)\rangle$ are given by [59]

$$\langle\sigma_z(\infty)\rangle_{k,\text{metric}} = \frac{(2k^2-\gamma^2)e^{-2\pi\delta}-k^2}{k^2-\gamma^2 e^{-2\pi\delta}}$$

$$\langle\sigma_z(\infty)\rangle_{k,\text{norm}} = \frac{2ke^{-2\pi\delta}-k+\gamma}{2\gamma e^{-2\pi\delta}+k-\gamma} \quad (6)$$

where the different regimes of non-Hermiticity are dictated by the magnitude of $(\gamma^2/F)$. In the Hermitian limit $\gamma \to 0$, both $\langle\sigma_z(\infty)\rangle_{k,\text{metric}}$ and $\langle\sigma_z(\infty)\rangle_{k,\text{norm}}$ converge to the standard Landau-Zener result $2e^{-2\pi\delta_0}-1$ where $\delta_0 = (k^2/2F)$ [58]. It is worth noting that the discrepancy observed between the computations of the operator $\sigma_z$ in Eq. (6) is also observed in the expectation value of other operators, such as $\sigma_x$, whose expectation value displays no definite symmetry in the norm computation scheme, but presents an odd symmetry within the metric framework. To summarize, in general, the metric dramatically alters the dynamics of the non-Hermitian system and correctly reflects the symmetry structures of the observables [59].

*Adiabatic limit.*—We now turn to the adiabatic limit $F \to 0$. For $\gamma = 0$, a universal KZ scaling of defects emerges in the adiabatic limit. For the non-Hermitian case where $\gamma \neq 0$, we first remark that the adiabatic limit corresponds to the regime of strong non-Hermiticity $(\gamma^2/F) \to \infty$ in our model. The presence or absence of the aforementioned symmetry in physical observables, as obtained from the metric vs the normalization methods, leads to a direct physical consequence in this limit.

In analogy to the Hermitian Landau-Zener and KZ problem, the defects are defined as the excitations which move away from the south pole of the Bloch sphere. Note that the south pole of the Bloch sphere corresponds to the ground state of the Hermitian end point. The density of defects is then given by [27]

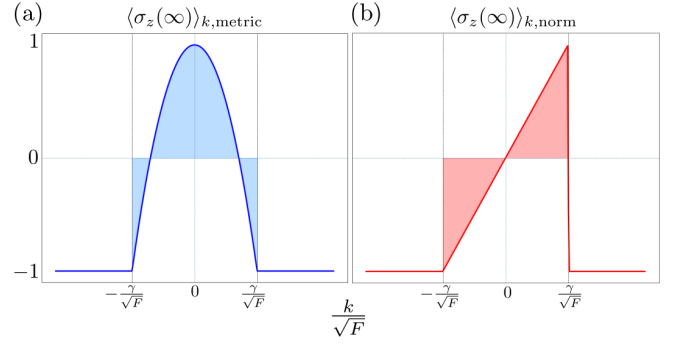

FIG. 3. The asymptotic value of the spin expectation values, given by Eq. (6), in the adiabatic limit $F \to 0$ (here $(\gamma^2/F) = 400$ and $\gamma = 1$). The shaded areas show the defect contribution from the $\mathcal{PT}$-broken modes. The behavior of the $\mathcal{PT}$-symmetric modes is accurately captured by both methods, as shown in (a) and (b). However, the effect of defect freezing is only captured when the metric is taken into account, as shown in (a). In contrast, the shaded areas cancel out in (b). This is a direct consequence of the odd parity of $\langle\sigma_z(\infty)\rangle_{k,\text{norm}}$ with respect to $k$.

$$\Sigma_z = \Sigma_z^{\mathcal{PT}s} + \Sigma_z^{\mathcal{PT}b}$$

$$\Sigma_z^{\mathcal{PT}s/b} = \int_{k\in\mathcal{PT}s/b}\frac{dk}{2\pi}\lim_{F\to 0}\langle\sigma_z(\infty)\rangle_k \quad (7)$$

where $\mathcal{PT}s$ and $\mathcal{PT}b$ indicate the contributions from the modes undergoing $\mathcal{PT}$-symmetric and $\mathcal{PT}$-broken evolution, $|k| \geq |\gamma|$ and $|k| < |\gamma|$, respectively. The asymptotic expression $\langle\sigma_z(\infty)\rangle_k$ is given by Eq. (6). For the $\mathcal{PT}$-broken modes, the metric and the norm methods predict starkly different asymptotic behaviors in the adiabatic limit. We obtain $\langle\sigma_z(\infty)\rangle_{k,\text{metric}} \to 1 - (2k^2/\gamma^2)$, which preserves the even parity in $k$, consistent with the symmetry analysis [59]. On the other hand, the normalization approach yields $\langle\sigma_z(\infty)\rangle_{k,\text{norm}} \to (k/\gamma)$ which is odd in $k$, in contrast to Eq. (6) which has no definite parity in $k$. This shows that using the normalization approach, a definite parity in the $\sigma_z$ observable only emerges in the adiabatic limit. This is shown in Fig. 3. The contribution of the $\mathcal{PT}$-broken modes to the defect density is thus

$$\left(\Sigma_z^{\mathcal{PT}b}\right)_{\text{metric}} = \frac{\gamma}{3\pi}$$
$$\left(\Sigma_z^{\mathcal{PT}b}\right)_{\text{norm}} = 0. \quad (8)$$

The nonzero defect contribution from the $\mathcal{PT}$-broken modes shows that defects are generated when this system is driven across an exceptional point, no matter how slow the drive is. This shows the violation of quantum adiabaticity, consistent with the fact that non-Hermitian systems are inherently out of equilibrium. This is in stark contrast to the Hermitian case where the defect density tends to zero as $F \to 0$ [58], and is corroborated by the findings of recent experimental work [46]. However, this defect freezing effect, where one does not recover the state of the final state Hamiltonian in the adiabatic limit, is not captured if we do not take the dynamics of the metric into account. This is a direct consequence of the odd parity of $\langle\sigma_z(\infty)\rangle_{k,\text{norm}}$ with respect to $k$ in the adiabatic limit where $F \to 0$.

We saw in Fig. 2(b) that, away from the adiabatic limit, the time-evolved state $|\Psi(t)\rangle_k$ shows a nontrivial behavior even for $\mathcal{PT}$-symmetric modes. However, a clear distinction in the behaviors between $\mathcal{PT}$-symmetric and $\mathcal{PT}$-broken modes is recovered in the adiabatic limit. This is shown in Fig. 3. For the $\mathcal{PT}$-symmetric modes, the metric and the norm methods predict the same asymptotic behaviors: $\langle\sigma_z(\infty)\rangle_k \to -1$ and thus $\Sigma_z^{\mathcal{PT}s} = (\gamma/\pi) - 1$. In this limit, the $\mathcal{PT}$-symmetric modes are pinned to the south pole of the Bloch sphere, where the term $(\gamma/\pi)$ in $\Sigma_z^{\mathcal{PT}s}$ shows a reduction in the fraction of spins pointing to the south pole compared to the Hermitian case. We emphasize that these are not the defects.

Conventionally, Hermitian many-body systems are expected to display a power-law scaling of the defects generated after a slow ramp across a critical point. For a generic spin system, this would mean $\sigma_z = -1 + \mathcal{O}(F^\theta)$ leading to a defect density $\sim F^\theta$, where $\theta$ depends on the critical exponents at equilibrium. For an infinite ensemble of Hermitian two-level systems, one has $\theta = \frac{1}{2}$ [44,64,65]. Interestingly, while the metric approach maps the non-Hermitian problem to an effectively Hermitian one, it captures the defect freezing effect that stems from the truly nonequilibrium nature of our non-Hermitian Hamiltonian. Further corrections to the defect density may still obey the KZ mechanism, but possibly with a different power law than shown in Ref. [27]. It is worth noting that, while the rate-independent result in Eq. (8) is rather remarkable for an effectively Hermitian ensemble of two-level systems, a similar violation of KZ scaling has already been observed when crossing infinitely degenerate critical points [66–68].

*Conclusion.*—Our work shows that quantum adiabaticity is violated in our non-Hermitian system, as defects are created purely by the $\mathcal{PT}$-broken modes, which survive even in the adiabatic quench limit. This is consistent with the spectral coalescence at the EPs leading to ambiguity across a quench. The normalization approach completely misses this fundamental feature, which is a direct consequence of the symmetry structure of the calculated observable. Our results can be experimentally verified in a variety of quantum-engineered systems where non-Hermitian drives can be directly implemented. For example, the evolution of the metric can be directly engineered using quantum gates [17–19], single-photon interferometry [45], and parametric amplification [69]. Many open questions regarding the dynamics of non-Hermitian systems remain, such as the postquench spread of correlation and the putative violation of Lieb-Robinson bounds [6,26,30].

This research was funded by the Swiss National Science Foundation (SNSF) Grant No. 200021 207537 and by the Deutsche Forschungsgemeinschaft (DFG, German Research Foundation) under Germany's Excellence Strategy EXC2181/1-390900948 (the Heidelberg STRUCTURES Excellence Cluster) and a Simons Investigator Award. The authors would like to thank G. M. Graf for numerous fruitful discussions and E. Bergholtz for comments on our manuscript.